\documentclass{Interspeech}

\usepackage{amsmath,graphicx}
\usepackage{cite}
\usepackage{colortbl}
\usepackage{comment}
\usepackage{adjustbox}
\usepackage{booktabs}
\usepackage{arydshln}
\usepackage{multirow}

\usepackage{amssymb}
\usepackage{pifont}

\usepackage{graphicx}
\usepackage{amsmath}
\usepackage{amssymb}
\usepackage{booktabs}
\usepackage{algorithm}
\usepackage{listings}
\usepackage{placeins}
\usepackage[accsupp]{axessibility}
\usepackage{nccmath}
\usepackage{tcolorbox}

\usepackage{caption} 
\usepackage{subcaption}
\usepackage{makecell}

\definecolor{ftcolor}{rgb}{1,1,1} 
\definecolor{baselinecolor}{rgb}{1,1,1}
\definecolor{contrcolor}{rgb}{1.0, 0.98, 0.8}
\definecolor{predcolor}{rgb}{0.74, 0.83, 0.9}
\definecolor{decorrcolor}{rgb}{0.98, 0.85, 0.87} 
\definecolor{knowcolor}{rgb}{0.80, 0.94, 0.75}
\definecolor{offlinecolor}{rgb}{1,1,1}
\definecolor{firsttaskcolor}{rgb}{0.97, 0.97, 0.97}
\definecolor{azure(colorwheel)}{rgb}{0.0, 0.5, 1.0}
\definecolor{gray(x11gray)}{rgb}{0.75, 0.75, 0.75}
\definecolor{lightgray}{rgb}{0.90, 0.90, 0.90}
\definecolor{darkgray}{rgb}{0.66, 0.66, 0.66}
\definecolor{teagreen}{rgb}{0.82, 0.94, 0.75}
\definecolor{almondlow}{RGB}{252,239,219} 
\definecolor{almondmiddle}{RGB}{237,225,206} 
\definecolor{almondhigh}{RGB}{224,213,194} 
\definecolor{almondultra}{RGB}{214,204,186} 
\definecolor{greylow}{RGB}{235,235,235} 
\definecolor{greymiddle}{RGB}{211,211,211} 
\definecolor{greyhigh}{RGB}{192,192,192} 
\definecolor{pinksecondbest}{RGB}{252, 241, 241}

\definecolor{yellow}{RGB}{253, 250, 206}
\definecolor{lightviolet}{RGB}{208,187,236}
\definecolor{pastelred}{RGB}{232, 131, 131}
\definecolor{pastelviolet}{rgb}{0.8, 0.7, 0.79}
\definecolor{champagne}{rgb}{0.97, 0.91, 0.81}
\definecolor{lightblue}{RGB}{225, 241, 255}

\NewDocumentCommand\emojice{}{
    \includegraphics[scale=0.007]{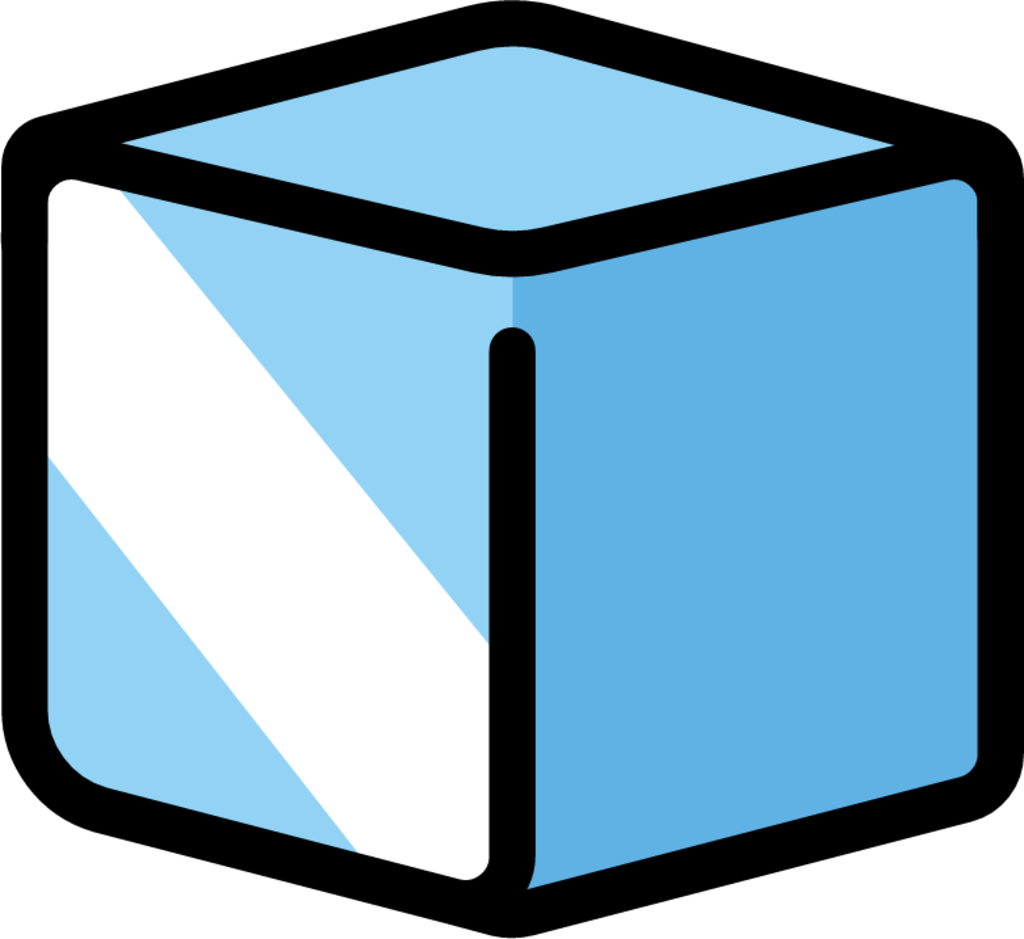}
}

\NewDocumentCommand\emojifire{}{
    \includegraphics[scale=0.007]{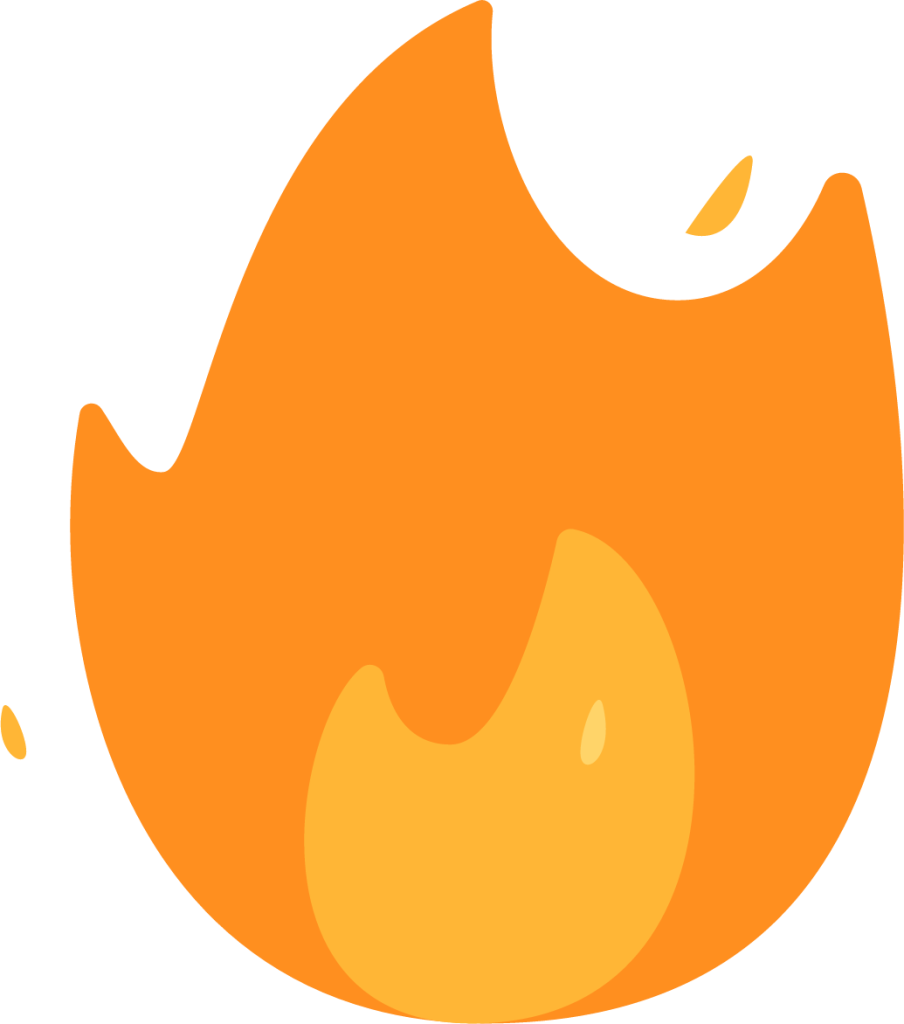}
}

\newcommand\blfootnote[1]{%
  \begingroup
  \renewcommand\thefootnote{}\footnote{#1}%
  \addtocounter{footnote}{-1}%
  \endgroup
}

\interspeechcameraready

\title{Scaling and Enhancing LLM-based AVSR: \\ A Sparse Mixture of Projectors Approach}

\author[affiliation={1}]{Umberto}{Cappellazzo}
\author[affiliation={2}]{Minsu}{Kim}
\author[affiliation={1}]{Stavros}{Petridis}
\author[affiliation={3}]{Daniele}{Falavigna}
\author[affiliation={3}]{Alessio}{Brutti}

\affiliation{}{Imperial College London}{UK}
\affiliation{}{Meta AI}{UK}
\affiliation{}{Fondazione Bruno Kessler}{Italy}
\email{u.cappellazzo@imperial.ac.uk}


\usepackage{comment}

\begin{document}

\maketitle

\begin{abstract}
Audio-Visual Speech Recognition (AVSR) enhances robustness in noisy environments by integrating visual cues. While recent advances integrate Large Language Models (LLMs) into AVSR, their high computational cost hinders deployment in resource-constrained settings. To address this, we propose Llama-SMoP, an efficient Multimodal LLM that employs a Sparse Mixture of Projectors (SMoP) module to scale model capacity without increasing inference costs. By incorporating sparsely-gated mixture-of-experts (MoE) projectors, Llama-SMoP enables the use of smaller LLMs while maintaining strong performance. We explore three SMoP configurations and show that Llama-SMoP \texttt{DEDR} (Disjoint-Experts, Disjoint-Routers), which uses modality-specific routers and experts, achieves superior performance on ASR, VSR, and AVSR tasks. Ablation studies confirm its effectiveness in expert activation, scalability, and noise robustness. 
\end{abstract}

\section{Introduction}
\label{sec:introduction}
\blfootnote{Only non-Meta authors conducted any of the dataset preprocessing (no dataset pre-processing took place on Meta’s servers or facilities).}
Automated speech recognition technologies have made significant progress and are widely employed in various real-world applications~\cite{zhang2023google}. In particular, Auditory Speech Recognition (ASR) technology~\cite{amodei2016deep,prabhavalkar2023end}, which uses audio as its input modality, is the most widely used and recognized by users. However, in real-world scenarios, audio can be corrupted by various background noises (e.g., speech captured in a crowded restaurant), leading recent research to focus on improving ASR robustness~\cite{dupont2000audio,narayanan2014investigation}. One key approach is leveraging multimodal inputs by integrating both audio and visual modalities. This technology, known as Audio-Visual Speech Recognition (AVSR)~\cite{noda2015audio, afouras2018deep, petridis2018audio, ma2021end, hong2022visual, ma2023auto, rouditchenko2024whisper}, achieves robust recognition even when input audio is severely corrupted with background noise by leveraging the correlation between audio and visual cues.

One chief research direction focuses on pre-training both audio and visual encoders through Self-Supervised Learning (SSL)~\cite{gui2024survey} to capture the intrinsic correlation between audio-visual modalities. This approach has demonstrated impressive performance even with limited labeled data (e.g., 30 hours), as reported in prior studies~\cite{shi2022learning, haliassos2023jointly, hsu2022u, haliassos2024braven, haliassos2024unified}. Building on this foundation, recent advances in generative AI and Large Language Models (LLMs)~\cite{achiam2023gpt,touvron2023llama, liu2024improved} have given rise to a new research line: aligning speech representations with LLMs~\cite{lakhotia2021generative,huang2024audiogpt,park2024lets,lu2024developing,tan2024ssr}. Notably, several studies have successfully adapted LLMs for automated speech recognition, including~\cite{chen2024s, hu2024large, ma2024embarrassingly, yu2024connecting, fathullah2024prompting, yeo2024where, Llama-AVSR}. However, despite their impressive performance, LLM-based speech recognition systems face a significant challenge: they require a large number of parameters. Recent works, such as Llama-AVSR~\cite{Llama-AVSR} and Llama-MTSK \cite{cappellazzo2025adaptive}, have shown that larger LLMs generally achieve better speech recognition performance, which is why previous works tend to use LLMs with over 7 billion parameters. This poses significant challenges, as these large-scale models cannot be easily deployed on parameter-constrained settings.

\begin{figure}[t]
    \centering
    \includegraphics[width=6cm]{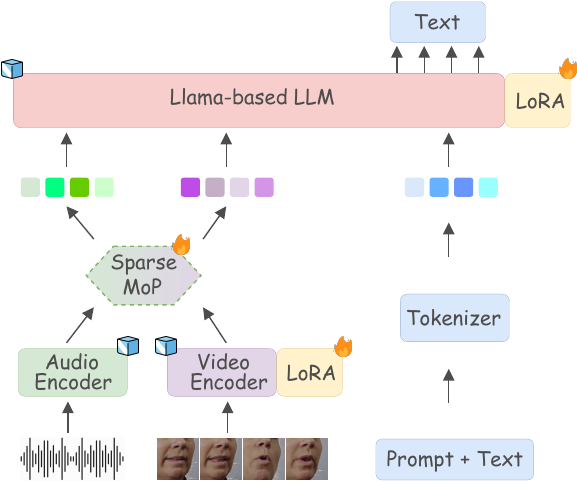}
    \caption{Illustration of the overall framework of the proposed Llama-SMoP model, where audio and video tokens are embedded using a sparsely-gated mixture-of-experts scheme.\emojifire and \emojice represent whether the module is trained or kept frozen.}
    \label{fig:main_diagram}
    \vspace{-0.6cm}
\end{figure}

In this paper, we investigate LLM-based AVSR systems, particularly focusing on the use of smaller LLMs (i.e., 1B and 3B parameters). We propose leveraging the Mixture of Experts (MoE) paradigm \cite{shazeer2017outrageously, lepikhin2020gshard, cappellazzo2024efficient, liu2024deepseek, he2024mixture, muennighoff2024olmoe} within the audio-visual projector to enhance the model capacity while keeping inference costs low. Recently, the integration of MoE blocks into Multimodal LLMs (MLLMs) has gained attention as a strategy for scaling encoders and LLMs. However, in LLM-based AVSR models, pre-trained encoders and LLMs are typically frozen during training \cite{Llama-AVSR}. To address this, we propose scaling the projector, an MLP layer, aligning with the trend of transforming LLMs' MLP layers in MoE-based models. While MoE has been applied to projectors in vision-language tasks \cite{li2024cumo} and chart understanding \cite{xu2024chartmoe}, these methods primarily emphasize single-modality inputs and efficient initialization through co-upcycling or task-specific alignment techniques.

\begin{figure*}[t]
    \centering
    \includegraphics[width=0.9\textwidth]{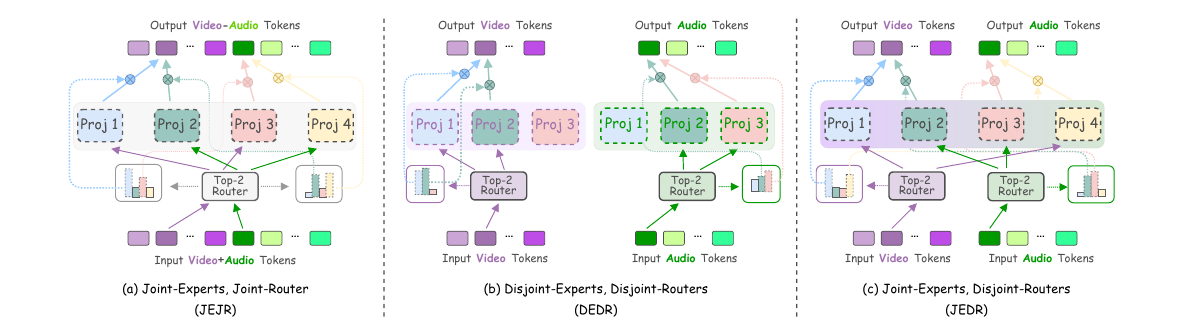}
    \vspace{-0.2cm}
    \caption{Detailed illustration of the three proposed SMoP configurations. (a) Joint-Experts, Joint-Router (\texttt{JEJR}) uses one multimodal router and one pool of expert for embedding audio-visual representations. (b) Disjoint-Experts, Disjoint-Routers (\texttt{DEDR}) uses modality-specific routers and experts for embedding modality-specific representations. (c) Joint-Experts, Disjoint-Routers (\texttt{JEDR}) uses modality-specific routers and one shared group of experts, so routers assign the Top-K experts considering the modality characteristics.}
    \label{fig:SMoP}
    \vspace{-0.6cm}
\end{figure*}

In this work, we introduce a novel module called \textbf{Sparse Mixture of Projectors} (SMoP) for embedding multimodal speech representations into the LLM space. Specifically, we investigate three different SMoP variants for processing the audio-visual tokens, based on different router and expert projector configurations. We concentrate on audio-visual models that utilize small-scale pre-trained encoders and LLMs, which typically exhibit lower performance compared to their large-scale counterparts. SMoP is particularly advantageous in these resource-constrained settings, as it improves performance using a small number of experts while incurring negligible additional parameter activation and computational overhead in inference.

We summarize the contributions of this work as follows:
\begin{itemize}
    \item We propose Llama-SMoP, an MLLM employing Sparse MoE to enhance audio-visual capabilities. The SMoP module is simple, efficient, and model-agnostic, allowing seamless integration with various pre-trained encoders and LLMs.
    \item Among the three proposed configurations, Llama-SMoP \texttt{DEDR} (Disjoint-Experts, Disjoint-Routers) achieves the best performance, outperforming previous methods across different-sized Llama-based LLMs on the AVSR task. We also demonstrate that Llama-SMoP is effective for both ASR and VSR tasks.
    \item We conduct ablation studies on expert activation frequency and optimal expert configurations, as well as showing that Llama-SMoP remains robust also in noisy scenarios.
\end{itemize}

\section{Llama-SMoP}
\label{sec:methodology}

We propose Llama-SMoP, an MLLM employing sparsely-gated mixture-of-experts~\cite{shazeer2017outrageously, lepikhin2020gshard} to increase model capacity without a proportional increase in computational cost. 
This is crucial in resource-constrained LLM-based AVSR systems, as we aim to improve performance despite using smaller-scale LLMs and pre-trained encoders. 
Llama-SMoP computes audio and video tokens via modality-specific pre-trained encoders, which are then fed to the LLM as prefix tokens (together with the textual tokens). This approach, denoted as decoder-only, is adopted by several architectures due to its versatility \cite{yao2024dense, liu2024visual, lin2024vila, fini2024multimodal, lee2024meteor}. Llama-SMoP consists of three main components: \textbf{1)} pre-trained audio and video encoders, \textbf{2)} a SMoP module, and \textbf{3)} an LLM parameter-efficiently fine-tuned via LoRA \cite{hu2021lora}. 

\textbf{Audio/Video Pre-Trained Encoders}. We use pre-trained audio and video encoders to project the input audio and video data into audio tokens $\mathbf{X}^{\mathsf{A}}$ and video tokens $\mathbf{X}^{\mathsf{V}}$. The pre-trained encoders are maintained \textit{frozen} during the training stage.

\textbf{SMoP Module}. The SMoP module replaces the standard projector with sparsely-gated MoE projectors \cite{shazeer2017outrageously}, each consisting of a two-layer MLP. Since we handle both audio and video modalities, we propose three router design strategies for processing the multimodal tokens. Before describing them in detail, we define how a generic SMoP module works. 

To scale up the model with multiple projectors in parallel, a SMoP module encompasses a router network $\mathcal{R}$ which chooses the top-K expert projectors out of the total $\mathsf{N}$ experts $\{E_i\}_{i=1}^{\mathsf{N}}$, and thus learns the optimal distribution over the experts for each token. For a given token $\mathbf{x}$, the router $\mathcal{R}$ picks the top-K experts based on the highest scores obtained using a learnable gating function (in our case a linear layer parameterized by $\mathbf{W}$), which are normalized via Softmax. The final output $\mathbf{z}$ is the linearly weighted combination of each expert’s output scaled by the corresponding gate's output as follows:
\begin{align}
\mathbf{z} &= \sum_{i=1}^\mathsf{N}\mathcal{R}(\mathbf{x})_i \cdot E_i(\mathbf{x}), \label{eq:1}
\\
 \mathcal{R}(\mathbf{x}) &=  \text{Top-K}(\texttt{Softmax}(\mathbf{x} \cdot  \mathbf{W}), \text{K}), \label{eq:2}
 \\
 \text{Top-K}(\mathbf{h}, \text{K}) &= \begin{cases} 
    \mathbf{h},  & \text{if $\mathbf{h}$ is in the Top-K},\\ 
    0, & \text{otherwise}.
  \end{cases} 
\end{align}
The SMoP module processes the audio $\mathbf{X}^{\mathsf{A}}$ and video $\mathbf{X}^{\mathsf{V}}$ tokens, obtaining the  output audio-visual tokens $\mathbf{Z}^{\mathsf{AV}}$.

We investigate \underline{three} different SMoP methods based on how we process the audio and video tokens, as shown in Fig.~\ref{fig:SMoP}. The first approach, denoted \textit{Joint-Experts, Joint-Router} (\texttt{JEJR}; Fig.~\ref{fig:SMoP}a), defines a \textit{single} \textit{joint} audio-visual router $\mathcal{R}^{\texttt{AV}}$ that dispatches the concatenation of audio and video tokens to a \textit{shared} pool of $\mathsf{N}$ audio-visual experts $\{E_i^{\texttt{AV}}\}_{i=1}^{\mathsf{N}}$. This method combines audio and video tokens into a single representation, allowing the model to learn cross-modal interactions directly. While \texttt{JEJR} can be effective for capturing these cross-modal interactions, it may lose some modality-specific details. 

In contraposition to \texttt{JEJR}, we explore the \textit{Disjoint-Experts, Disjoint-Routers} (\texttt{DEDR}; Fig.~\ref{fig:SMoP}b) configuration. It utilizes \textit{modality-specific} routers, $\mathcal{R}^{\texttt{A}}$ for audio and $\mathcal{R}^{\texttt{V}}$ for video, to direct each modality's tokens to its \textit{own dedicated} set of experts ($\{E_i^{\texttt{A}}\}_{i=1}^{\mathsf{N_A}}$ and $\{E_i^{\texttt{V}}\}_{i=1}^{\mathsf{N_V}}$, respectively). Output audio and video tokens are then concatenated before being processed by the LLM. This modular approach enables each expert pool to specialize in processing a specific modality, thereby maximizing the extraction and utilization of modality-specific representations. However, because modalities are processed separately, this approach may hinder the model's ability to learn cross-modal interactions. Finally, we also study a hybrid approach between the two previous ones: \textit{Joint-Experts, Disjoint-Routers} configuration (\texttt{JEDR}; Fig.~\ref{fig:SMoP}c). \textit{Modality-specific} routers, $\mathcal{R}^{\texttt{A}}$ and $\mathcal{R}^{\texttt{V}}$, assign each modality's tokens to a \textit{shared} pool of experts $\{E_i^{\texttt{AV}}\}_{i=1}^{\mathsf{N}}$. This setup allows the model to leverage a common pool of expertise while maintaining the distinctiveness of each modality during routing.

For the \texttt{JEJR} and \texttt{JEDR} configurations, we must guarantee that audio and video tokens have the same token hidden dimension because they are processed by the same group of experts. If the hidden dimension of audio and video encoders does not match, we add a linear layer to compensate for this. 

\textbf{LLM}. The role of LLM in our work is generating the corresponding speech-recognized transcription  $\mathbf{Y} = \{y_l\}_{l=1}^{L}$ in an auto-regressive manner, conditioned on the input audio-visual tokens $\mathbf{Z}^{\mathsf{AV}}$ and prompt textual tokens $\mathbf{X}^P$, where $L$ represents the number of tokens of the ground truth transcription. Accordingly, the probability of the target $\mathbf{Y}$ is computed by:
\begin{equation}
\setlength{\abovedisplayskip}{3pt}
\setlength{\belowdisplayskip}{4pt}
p(\mathbf{Y}|\mathbf{Z}^{\mathsf{AV}}, \mathbf{X}^P) = \prod_{l=1}^{L}p(y_l|\mathbf{Z}^{\mathsf{AV}}, \mathbf{X}^P, y_{<l}),
\end{equation}
where $y_{<l}$ is the previous generated output sequence.

\textbf{Total Loss}. Besides the next-token prediction loss of the LLM, we include two auxiliary losses, which are commonly used
to avoid the router to activate only a few experts, dispensing with expert imbalance issues. The auxiliary losses comprise the load balancing loss $\mathcal{L}_{\text{b}}$ \cite{shazeer2017outrageously}, which penalizes unequal assignment of the experts, and the router z-loss $\mathcal{L}_{\text{z}}$ \cite{zoph2022st}, which penalizes large logits in the router that may cause instabilities. Finally, the total loss is defined as:
\begin{equation}
\setlength{\abovedisplayskip}{3pt}
\setlength{\belowdisplayskip}{4pt}
    \mathcal{L} = - \log p(\mathbf{Y}|\mathbf{Z}^{\mathsf{AV}}, \mathbf{X}^P) + \alpha_{\text{b}}\mathcal{L}_{\text{b}} + \alpha_{\text{z}}\mathcal{L}_{\text{z}},
\end{equation}
with $\alpha_{\text{b}}$ and $\alpha_{\text{z}}$ set to $0.01$ and $0.001$ following \cite{zoph2022st, shazeer2017outrageously, muennighoff2024olmoe}.

\section{Experiments and Results}
\label{sec:experiments}

\begin{table}
\centering
\caption{ASVR results of our three proposed SMoP configurations and baselines on LRS3. SMoP-\textbf{Y} means that each pool of experts contains \textbf{Y} experts. \textbf{Bold} and
\underline{underscore} numbers denote the best and second best scores, respectively.}
\vspace{-0.2cm}
\begin{tabular}{lccc}
    \toprule
      \multirow{2}{*}{\textbf{Method}} & 
      \multicolumn{3}{c}{\textbf{Llama Model}} \\
\cmidrule(l){2-4} & \cellcolor{lightviolet}\textbf{3.2-1B}&\cellcolor{teagreen} \textbf{3.2-3B} &\cellcolor{yellow} \textbf{3.1-8B} \\
    \midrule
     Llama-AVSR \cite{Llama-AVSR}& 3.81 &2.80 & \underline{1.09}\\ \hdashline \addlinespace[2pt]
     DCI \cite{yao2024dense} & 3.46 &2.60 & 1.28\\
     MM-Fuser \cite{cao2024mmfuser} & 3.45 &2.66 & 1.31\\
     \hdashline \addlinespace[2pt]
     \textbf{SMoP-4} \texttt{JEJR} & 3.97 &2.63 & 1.53\\
     \textbf{SMoP-4} \texttt{JEDR} & 4.16 & 2.80 & 1.23\\
     \rowcolor{lightblue}\textbf{SMoP-3} \textbf{\texttt{DEDR}} & \textbf{3.31} &\textbf{2.29} & \textbf{0.96}\\
     \hdashline \addlinespace[2pt]
     \textbf{SMoP-4}-V + DCI-A & \underline{3.34} &3.05 & 2.79\\
     \textbf{SMoP-4}-V + MM-Fuser-A & 3.45 &\underline{2.51} & 1.86 \\
     
 \bottomrule
\end{tabular}
\label{tab:main}
\vspace{-0.5cm}
\end{table}
\subsection{Implementation Details}
\textbf{Datasets}. We train and evaluate Llama-SMoP on LRS3 \cite{afouras2018lrs3}, the largest publicly available dataset for AVSR. LRS3 contains $433$ hours of transcribed English video clips from TED talks.

\textbf{Pre-Processing}. We follow \cite{ma2023auto, Llama-AVSR} for the pre-processing of the dataset. For the video modality, we crop the mouth region of interests (ROIs) through a bounding box of $96$ × $96$. Each frame is normalised by subtracting the mean and dividing by the standard deviation of the training set. Audio data is preprocessed with z-normalisation per utterance.

\textbf{Tasks}. 
The AVSR task is studied for the main results, and we also report the results for the ASR/VSR tasks in Section~\ref{subsec:ablations}.

\textbf{Llama-SMoP Details}. We use multiple Whisper models \cite{radford2023robust} (Tiny, Base, Small, Medium) as pre-trained audio encoders, whilst AV-HuBERT Large \cite{shi2022learning} is used for computing the video tokens for all experiments. Their weights remain frozen throughout the training phase, and only for the VSR task do we equip the video encoder with LoRA modules following \cite{Llama-AVSR}. Each expert projector is a two-layer MLP . We use $4$ shared experts for the \texttt{JEJR} and \texttt{JEDR} configurations (e.g., \textbf{SMoP-4} \texttt{JEJR} in Table~\ref{tab:main}), while $3$ experts for each modality-based SMoP module in \texttt{DEDR}, resulting in $6$ experts overall. We use Token-Choice Top-K as routing strategy \cite{shazeer2017outrageously, lepikhin2020gshard, muennighoff2024olmoe}, where we select and activate the top-K experts for each input token, with K = $2$. As for the LLM, we experiment with $3$ pre-trained models from the Llama 3 family of varying size \cite{dubey2024llama}: Llama 3.2-1B, Llama 3.2-3B, and also Llama 3.1-8B to confirm the effectiveness of SMoP under a large-scale parameter setting. As in \cite{Llama-AVSR}, we parameter-efficiently fine-tune the LLM via LoRA \cite{hu2021lora, he2021towards, cappellazzo2024parameter}. Following \cite{Llama-AVSR, fathullah2024prompting, ma2024embarrassingly, fang2024llama}, we reduce the number of tokens processed by the LLM by stacking multiple consecutive tokens along the hidden dimension. We apply a compression rate of $3$ both for audio and video tokens.

\textbf{Baselines}. We compare Llama-SMoP with Llama-AVSR \cite{Llama-AVSR}, which serves as a baseline using a single projector without any scaling method. Additionally, we compare it with two recent scaling methods that integrate deep and shallow intermediate features from pre-trained encoders (in our case, audio and video encoders) by implementing them into Llama-AVSR. Specifically, Dense Channel Integration (DCI) \cite{yao2024dense} incorporates features from all layers before concatenating them with the final layer's features across the hidden dimension, whereas MM-Fuser \cite{cao2024mmfuser} fuses features from different layers using an attention-based module. Both DCI and MM-Fuser have been shown to enhance the visual representations of existing vision-language MLLMs. To the best of our knowledge, we are the first to investigate their efficacy in AVSR.

\textbf{Training/Inference Details}. Following \cite{Llama-AVSR, ma2023auto}, we augment visual inputs through horizontal flipping, random cropping, and adaptive time masking, while for audio we only apply adaptive time masking. We define the textual prompts as:  ``\texttt{Transcribe \{\textbf{task\_prompt}\} to text.}'', where \texttt{\textbf{task\_prompt}} $\in$ \{``\texttt{speech}'', ``\texttt{video}'', ``\texttt{speech and video}''\}. We train our model for $10$ epochs with the AdamW optimizer with cosine annealing scheduler and weight decay set to $0.1$ using NVIDIA A100 GPUs. The learning rate is set to 1e-3 for ASR and AVSR tasks, and 5e-4 for VSR. For decoding, we use beam search with a beam width of $15$ and temperature of $0.6$. The evaluation metric is the Word Error Rate (WER, \%). 

\begin{table}[t]
\renewcommand{\tabcolsep}{1.3mm}
\centering
    \caption{ASR/AVSR results under different acoustic noise levels.}
\vspace{-0.2cm}
\begin{tabular}{lcccccc}

\toprule
\multirow{2}{*}{\textbf{Method}} & \multirow{2}{*}{\textbf{Modality}} & \multicolumn{5}{c}{\cellcolor{almondlow}\textbf{SNR Level (dB)}}\\
  \cmidrule(rl){3-7} & & 7.5 & 5 & 2.5 & 0 & -2.5\\

\midrule
Llama-AVSR \cite{Llama-AVSR} & A & 6.3 & 10.0 & 19.1 & 35.1 & 95.1 \\ \hdashline \addlinespace[2pt]
DCI \cite{yao2024dense} & A & 6.2 & 10.1 & 18.7 & 33.3 & 93.0 \\
MM-Fuser \cite{cao2024mmfuser} & A & 5.8 & 9.3 & 17.6 & 33.0 & 89.4 \\
\rowcolor{lightblue} \textbf{SMoP-4} & A & 6.1 & 9.8 & 18.3 & 31.4 & 87.5 \\
\hline \addlinespace[2pt]
Llama-AVSR \cite{Llama-AVSR} & A-V & 5.2 & 7.1 & 10.4 & 11.3 & 27.5 \\ \hdashline \addlinespace[2pt]
DCI \cite{yao2024dense} & A-V & 4.9 & 6.8 & 9.6 & 9.8 & 23.1 \\
MM-Fuser \cite{cao2024mmfuser} & A-V & 4.7 & 6.1 & 9.3 & 9.8 & 24.5 \\
\rowcolor{lightblue} \textbf{SMoP-3} \textbf{\texttt{DEDR}} & A-V & 4.5 & 6.1 & 8.9 & 9.5 & 22.9 \\

\bottomrule
 \end{tabular}
\label{tab:noisy}
\vspace{-0.5cm}
\end{table}

\subsection{AVSR Performance Evaluation}

\begin{figure*}
\centering
\begin{subfigure}{0.31\textwidth}
    \includegraphics[width=\textwidth]{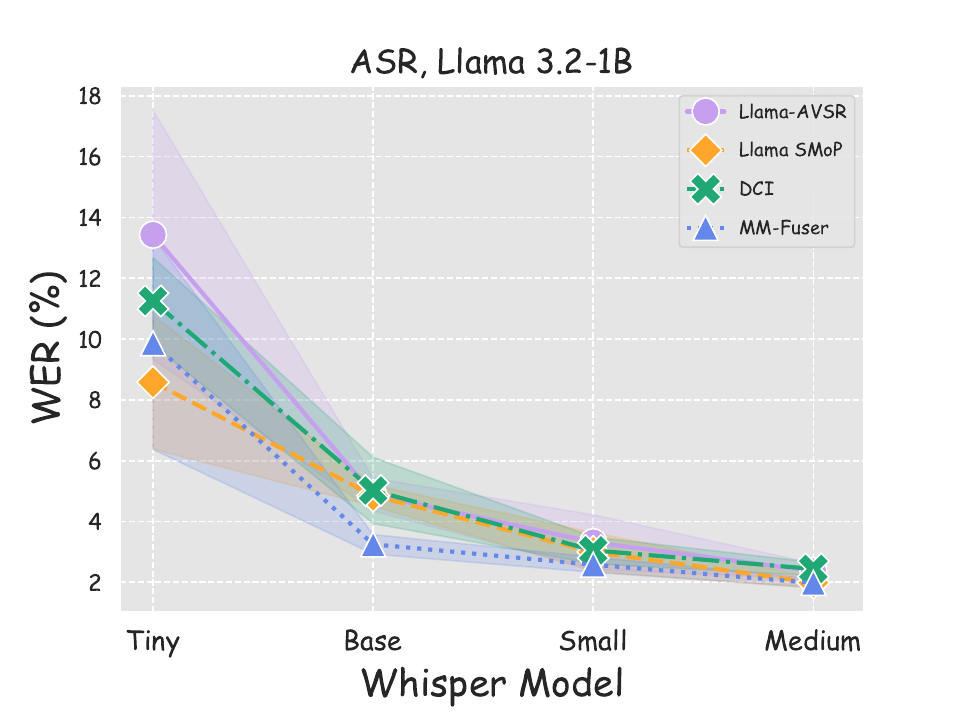}
\end{subfigure}
\begin{subfigure}{0.31\textwidth}
    \includegraphics[width=\textwidth]{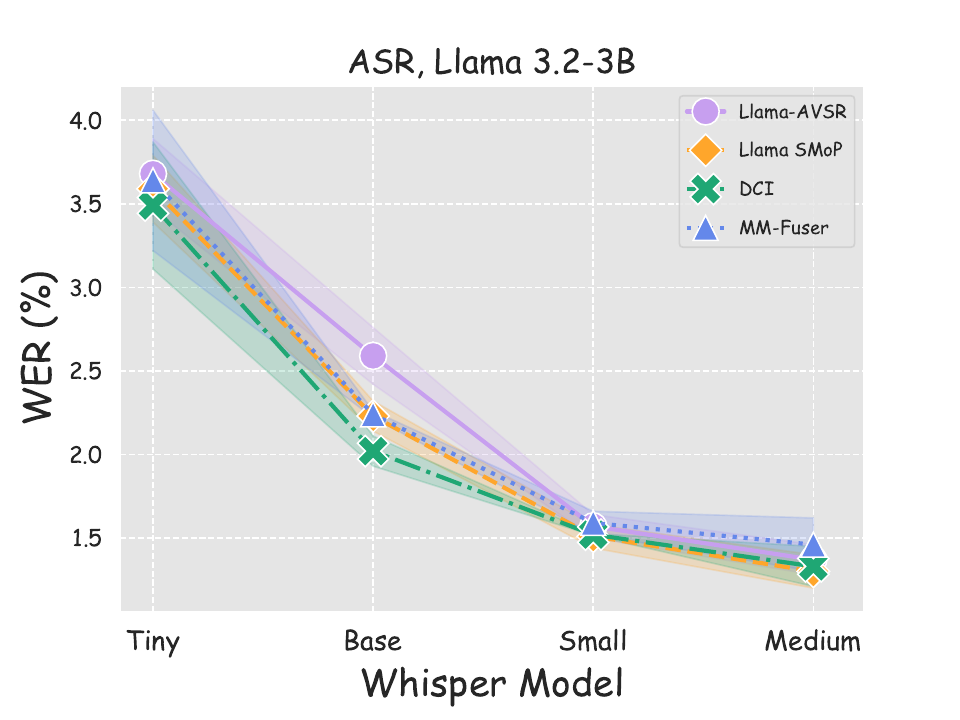}
    
\end{subfigure}
\begin{subfigure}{0.37\textwidth}
    \includegraphics[width=\textwidth]{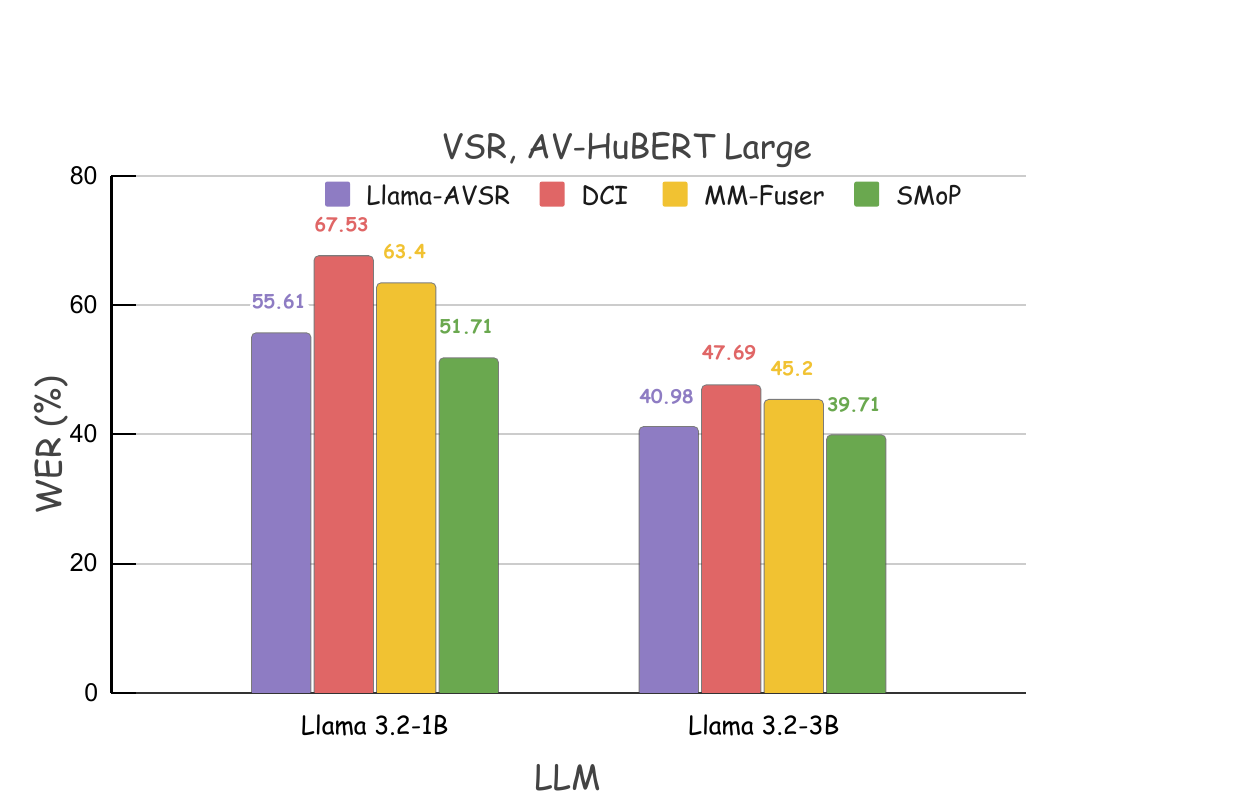}
\end{subfigure}
\caption{\textbf{(Left)}. ASR results for Llama-SMoP using different-size Whisper models with Llama 3.2-1B. \textbf{(Middle)}. ASR results for Llama-SMoP using different-size Whisper models with Llama 3.2-3B. \textbf{(Right)}. VSR results for Llama-SMoP with Llama 3.2-1B/3.2-3B.}
\label{fig:one_modality_ablation}
\vspace{-0.4cm}
\end{figure*}

In Table \ref{tab:main}, we compare AVSR results on LRS3 across our three SMoP strategies and the baseline models. Based on the ablation studies provided in Section~\ref{subsec:ablations} for ASR/VSR, we also report results for a hybrid approach that applies SMoP for video and DCI/MM-Fuser for audio. For Llama 3.2-1B and 3.2-3B, we use Whisper Base as the audio encoder, whereas for Llama 3.1-8B, we use Whisper Medium. Among the three SMoP variants, \texttt{DEDR} achieves the best results, improving the baseline Llama-AVSR across all LLM configurations and outperforming both DCI and MM-Fuser. SMoP-4 \texttt{JEJR} and \texttt{JEDR} provide little or no improvement over the Llama-AVSR baseline, suggesting that learning experts and routers independently is more effective. We also observe that our proposed SMoP module can be effectively combined with MM-Fuser and DCI, achieving improvements over Llama-AVSR. Finally, in the best configuration with Llama 3.1-8B and Whisper Medium (last column), neither exploiting intermediate features (i.e., DCI and MM-Fuser) nor using SMoP leads to improvements, except for SMoP-3 \texttt{DEDR}. This demonstrates that these methods are highly effective when using smaller encoders and LLMs, making them well-suited for resource-constrained scenarios.

\begin{figure}
\centering
    \includegraphics[width=6.9cm]{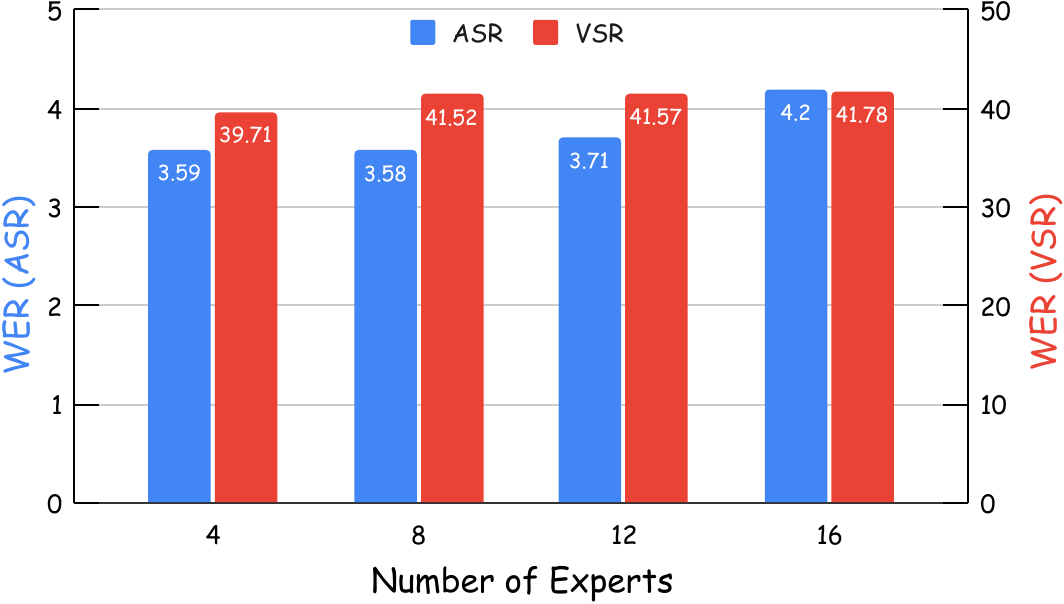}
\caption{Ablation analysis on the number of expert projectors for the ASR and VSR tasks.}
\label{fig:experts_number}
\vspace{-0.5cm}
\end{figure}
\subsection{Robustness against Noise} In Table \ref{tab:noisy}, we evaluate Llama-SMoP under noisy conditions. Following \cite{Llama-AVSR}, we inject babble noise from the NOISEX dataset at various SNR levels in inference. We report results for the ASR and AVSR task, using Whisper Base and Llama 3.2-3B. For both tasks, Llama-SMoP proves to be the most resilient method, particularly as the noise level increases. Overall, both Llama-SMoP and DCI/MM-Fuser outperform Llama-AVSR.

\subsection{Ablation on Model Parameter Scaling for ASR/VSR}
\label{subsec:ablations}
In this section, we study the ASR and VSR tasks by varying the sizes of the audio encoder and LLM. In this case, we use a single modality-specific router and a single pool of experts.

\textbf{ASR Results}. For ASR, we investigate Llama-SMoP and DCI/MM-Fuser with Whisper models of different sizes for Llama 3.2-1B (Fig.~\ref{fig:one_modality_ablation}, Left) and Llama 3.2-3B (Fig.~\ref{fig:one_modality_ablation}, Middle). We observe that the WER improvement achieved by these methods over the baseline Llama-AVSR diminishes as the size of the Whisper encoder and LLM increases. This suggests that these methods are particularly beneficial when the model size is smaller, aligning with the AVSR trend observed in Table~\ref{fig:main_diagram}.

\textbf{VSR Results}. For the VSR task, Llama-SMoP achieves a nearly 4-point reduction in WER when using Llama 3.2-1B, demonstrating superior performance (Fig.~\ref{fig:one_modality_ablation}, Right). In contrast, both DCI and MM-Fuser degrade performance in both configurations. We attribute this to the inherently higher WER in VSR, which may exacerbate the ambiguity of visual speech (i.e., lip movements) when integrating intermediate video features. 

\begin{figure}
\centering
    \includegraphics[width=6.9cm]{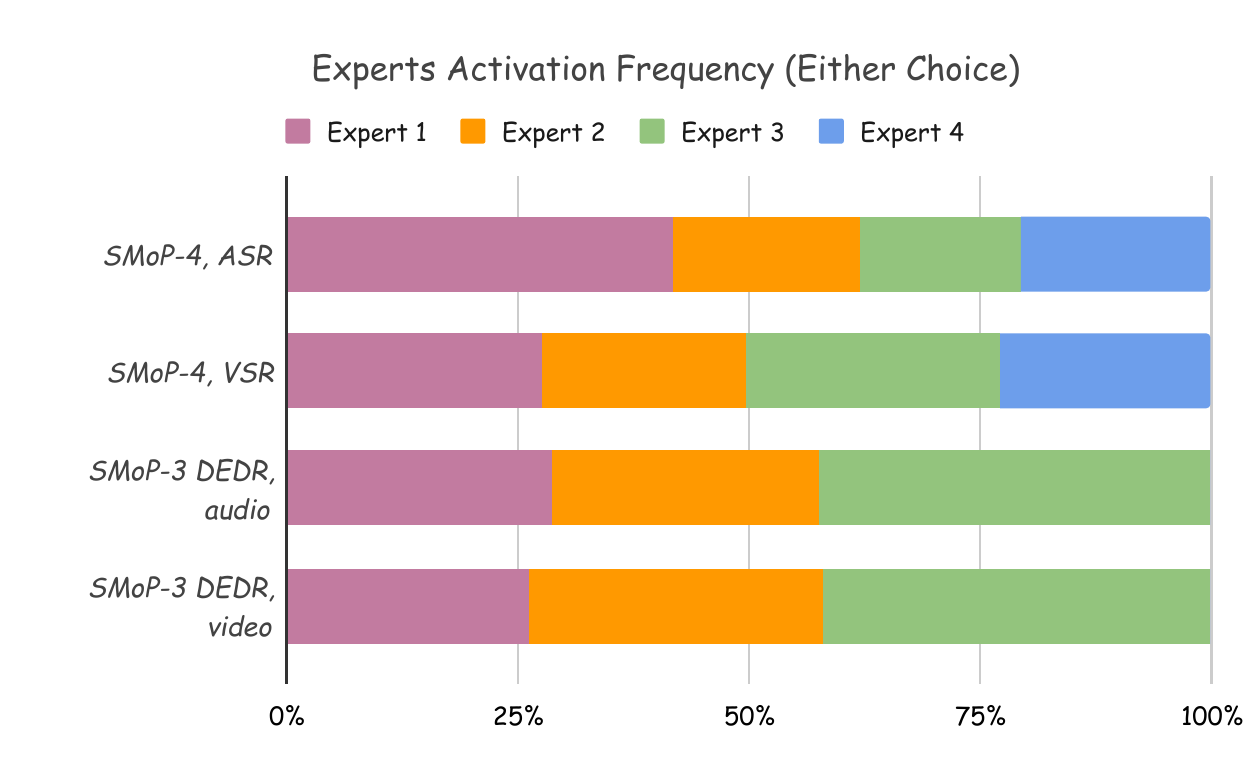}
\vspace{-0.2cm}
\caption{Proportion of tokens assigned to each expert, either as first or second choice.}
\label{fig:experts_activation}
\vspace{-0.5cm}
\end{figure}

\subsection{Optimal Expert Configuration for SMoP}
We study the optimal number of expert projectors for each SMoP module in ASR and VSR tasks, evaluating configurations with $4$, $8$, $12$, and $16$ experts. As shown in Fig.~\ref{fig:experts_number}, increasing the number of experts for ASR and VSR does not improve performance; instead, it slightly degrades it, likely due to redundant learning of similar tokens by additional experts. Recent MoE-based AVSR models also use $4$ \cite{cheng2024mixtures} or $8$ \cite{kim2025mohave} experts. Finally, Fig.~\ref{fig:experts_activation} presents the activation ratio of experts during inference, whether as the first or second choice. The router evenly activates the experts, ensuring all contribute to the computation of the output tokens. For ASR, one expert exhibits slightly higher activation, likely due to the relative simplicity of the task.

\section{Conclusion}
\label{sec:conclusion}

We present Llama-SMoP, an MLLM optimized for improved audio-visual processing. Its key innovation is replacing the linear projector with a Top-K sparse MoE module.  This approach allows for more efficient processing of multimodal audio-visual tokens, and we investigate three SMoP designs based on varying router and expert configurations. Llama-SMoP \texttt{DEDR} achieves superior WER results across multiple tasks, pre-trained encoders and LLMs. The SMoP module offers a simple and effective way to scale LLM-based AVSR under resource-constrained settings, incurring minimal additional inference overhead while boosting performance.  

{\bf Acknoledgment}: this work was partially funded by the European Union’s Horizon 2020 project ELOQUENCE (grant 101070558).

\bibliographystyle{IEEEtran}
\bibliography{mybib}

\end{document}